\documentclass[twocolumn,prd,aps,floats,showpacs]{revtex4}
\def\DESepsf(#1 width #2){\epsfxsize=#2 \epsfbox{#1}}
\usepackage{graphicx}
\usepackage{color}
\usepackage{amsmath}
\def\bmatrix{\left[\begin{array}}
\def\ematrix{\end{array}\right]}

\begin{document}

%

\let\a=\alpha      \let\b=\beta       \let\c=\chi        \let\d=\delta
\let\e=\varepsilon \let\f=\varphi     \let\g=\gamma      \let\h=\eta
\let\k=\kappa      \let\l=\lambda     \let\m=\mu
\let\o=\omega      \let\r=\varrho     \let\s=\sigma
\let\t=\tau        \let\th=\vartheta  \let\y=\upsilon    \let\x=\xi
\let\z=\zeta       \let\io=\iota      \let\vp=\varpi     \let\ro=\rho
\let\ph=\phi       \let\ep=\epsilon   \let\te=\theta
\let\n=\nu
\let\D=\Delta   \let\F=\Phi    \let\G=\Gamma  \let\L=\Lambda
\let\O=\Omega   \let\P=\Pi     \let\Ps=\Psi   \let\Si=\Sigma
\let\Th=\Theta  \let\X=\Xi     \let\Y=\Upsilon

%

%

\def\cA{{\cal A}}                \def\cB{{\cal B}}
\def\cC{{\cal C}}                \def\cD{{\cal D}}
\def\cE{{\cal E}}                \def\cF{{\cal F}}
\def\cG{{\cal G}}                \def\cH{{\cal H}}
\def\cI{{\cal I}}                \def\cJ{{\cal J}}
\def\cK{{\cal K}}                \def\cL{{\cal L}}
\def\cM{{\cal M}}                \def\cN{{\cal N}}
\def\cO{{\cal O}}                \def\cP{{\cal P}}
\def\cQ{{\cal Q}}                \def\cR{{\cal R}}
\def\cS{{\cal S}}                \def\cT{{\cal T}}
\def\cU{{\cal U}}                \def\cV{{\cal V}}
\def\cW{{\cal W}}                \def\cX{{\cal X}}
\def\cY{{\cal Y}}                \def\cZ{{\cal Z}}
%

\newcommand{\Ns}{N\hspace{-4.7mm}\not\hspace{2.7mm}}
\newcommand{\qs}{q\hspace{-3.7mm}\not\hspace{3.4mm}}
\newcommand{\ps}{p\hspace{-3.3mm}\not\hspace{1.2mm}}
\newcommand{\ks}{k\hspace{-3.3mm}\not\hspace{1.2mm}}
\newcommand{\des}{\partial\hspace{-4.mm}\not\hspace{2.5mm}}
\newcommand{\desco}{D\hspace{-4mm}\not\hspace{2mm}}
\renewcommand{\figurename}{Fig.}
\newcommand{\NTU}{Department of Physics, National Taiwan University,
    Taipei, Taiwan 10617}
\newcommand{\NCTSn}{National Center for Theoretical Sciences, North Branch,
    National Taiwan University, Taipei, Taiwan 10617}


%
\title{\boldmath
An Updated Numerical Analysis of eV Seesaw with Four Generations}
\vfill
\author{Wei-Shu Hou$^{1,2}$ and Fei-Fan Lee$^{1}$\\
{$^{1}$\NTU}\\
{$^{2}$\NCTSn}
 }

\date{\today}
%
%
%
\begin{abstract}
We consider the so-called ``eV seesaw'' scenario, with right-handed
Majorana mass $M_R$ at eV order, extended to four lepton generations.
The fourth generation gives a heavy pseudo-Dirac neutral lepton,
which largely decouples from other generations and is relatively stable.
The framework naturally gives 3 active and 3 sterile neutrinos.
We update a previous numerical analysis of a 3+3 study of the LSND anomaly,
taking into account the more recent results from the MiniBooNE experiment.
In particular, we study the implications for the third mixing angle
$\mathrm{sin}^2\theta_{13}$, as well as CP violation. We find that
current data do not seriously constrain more than one sterile neutrinos.
\end{abstract}
%
\maketitle

\pagestyle{plain}

\section{Introduction}

Neutrino data hint at the possibility of something beyond three massive,
mostly active neutrinos. The LSND result of $P(\bar{\nu}_\mu
\rightarrow \bar{\nu}_e)=0.264\%$~\cite{Lsnd1} can be explained
if there exist one sterile neutrino, with $\D m^2 \sim 1\ \rm{eV}^2$
with respect to other neutrinos. This lead to the so-called ``eV seesaw''
scenario~\cite{Gouvea}, where the right-handed neutrino Majorana
scale $M_R$ is taken to be of eV order. In 2006, the eV seesaw scenario was
extended to four lepton generations~\cite{Hou1}.
The fourth generation gives rise to a heavy pseudo-Dirac
neutral lepton, which largely decouples from the other
generations, and is relatively stable. The framework gives
naturally 3 active and 3 sterile neutrinos (3+3).

Motivated by the interpretation of the $\bar{\nu}_e$ excess
observed by LSND in terms of $\bar{\nu}_\mu \rightarrow
\bar{\nu}_e$ antineutrino oscillations, the MiniBooNE (MB)
experiment was constructed. In 2007, the MiniBooNE collaboration
presented~\cite{Miniboone1} their first results of a search for
$\nu_\mu \rightarrow \nu_e$ neutrino oscillations, which did not
support the LSND excess, but some excess at low energy was not
fully understood. Taking this result into account, Maltoni and
Schwetz~\cite{Maltoni1} found that the 3+2 scheme can fit both
LSND $\bar{\nu}_\mu \rightarrow \bar{\nu}_e$ data and MB $\nu_\mu
\rightarrow \nu_e$ data, and at the same time account for the
excess of low energy events in MB data. More recently, the
MiniBooNE collaboration reported their latest data, updating
neutrino results~\cite{Miniboone2} (including the low energy
region), as well as the first antineutrino
results~\cite{Miniboone3}, and first results from the off-axis
NuMI beam observed in the MiniBooNE detector~\cite{Adamson}.
Motivated by these new results, the authors of
Ref.~\cite{Karagiorgi} re-examined the 3+1 and 3+2 global fits to
the short-baseline (SBL) data. They found that the 3+2 oscillation
hypothesis provides only a marginally better description of all
SBL data over the 3+1 oscillation hypothesis. In addition, a 3+1
fit to all antineutrino SBL data (MiniBooNE($\bar{\nu}$), LSND,
KARMEN~\cite{KARMEN}, Bugey~\cite{Bugey}, and CHOOZ~\cite{CHOOZ})
can yield $86\%$ $\chi^{2}$-probability and high compatibility.

In this paper we update the previous numerical analysis of the 3+3
study of the LSND anomaly, taking into account MB data. In
particular, we consider the implications for the third mixing
angle, and CP violation.
The outline is as follows. In Sec. II we update the previous
numerical solutions by inputting the latest neutrino oscillation
parameters from global data. In Sec. III we add the CP violation phases
to the leptonic mixing matrix and investigate the effect of phase factors,
and give a short summary in Sec. IV.

\section{UPDATED NUMERICAL ANALYSIS}

\subsection{Updated Fit}

In Ref.~\cite{Maltoni1}, Maltoni and Schwetz discussed different
types of oscillation frameworks that include one or more sterile
neutrinos at the eV scale, by considering the global
short-baseline (SBL) experiments data which included the MiniBooNE
first result~\cite{Miniboone1}. The analysis built on an earlier
study which showed that MiniBooNE (2007), LSND, and the null
appearance experiments (KARMEN and NOMAD~\cite{NOMAD}) are
compatible under a 3+2 sterile neutrino oscillation scenario with
large CP violation. Recently, in light of recently published
results from the MiniBooNE
experiment~\cite{Miniboone2,Miniboone3,Adamson}, Karagiorgi {\it
et al.}~\cite{Karagiorgi} re-examined sterile neutrino oscillation
models. They found that, with the addition of the new MiniBooNE
data sets, a 3+2 oscillation hypothesis provides only a marginally
better description of all short-baseline data over a 3+1
oscillation hypothesis. On the other hand, fits to
antineutrino-only data sets, including appearance and
disappearance experiments, are found significantly more
compatible, even within a 3+1 oscillation scenario.
We shall therefore use the best-fit values for mass splittings and mixing angles
obtained from 3+1 fits to all antineutrino SBL data in Ref.~\cite{Karagiorgi},
and the corresponding projected value of the LSND probability
$P(\bar{\nu}_\mu \rightarrow \bar{\nu}_e)=0.2334\%$~\cite{Thomas}.

\begin{table}[t]
\caption{Best fit values, $1\s$ and  $3\s$  intervals for
three-flavor neutrino oscillation parameters taken from Ref.~\cite{GonzalezGarcia:2010er}.  }
\begin{center}
\begin{ruledtabular}
\begin{tabular}{cccc}
& BEST FIT & $1\s$ & $3\s$
\\ \hline \\
$\D m_{21}^2\,(10^{-5} \; {\rm eV}^2)$ & $7.59$  & $7.39-7.79$  & $6.90-8.20$    \\
$\D m_{31}^2\,(10^{-3} \; {\rm eV}^2)$ & $+2.47$  & $+(2.35-2.59)$  & $+(2.10-2.84)$      \\
$                                    $ & $-2.36$  & $-(2.29-2.43)$  & $-(2.00-2.72)$      \\
$\sin^2{\theta_{12}}$ & $0.32$  & $0.30-0.34$  & $0.28-0.37$    \\
$\sin^2{\theta_{23}}$ & $0.46$  & $0.41-0.53$  & $0.34-0.65$     \\
$\sin^2{\theta_{13}}$ & $0.014$  & $0.005-0.025$  & $\leq 0.049$      \\
\end{tabular}
\end{ruledtabular}
\end{center}
\end{table}

\begin{table}[b]
\caption{Seven inputs taken from Ref.~\cite{GonzalezGarcia:2010er,Thomas,Karagiorgi} and the
 best fit values from the minimization process.}
\begin{center}
\begin{ruledtabular}
\begin{tabular}{ccc}
 & INPUT  & 2010
\\ \hline \\
$P(\bar{\nu}_\mu \rightarrow \bar{\nu}_e)(\%)$ & $0.2334$  & $0.2090$      \\
$\D m_{21}^2\,(10^{-5} \; {\rm eV}^2)$ & $7.59$  & $7.59$      \\
$\D m_{31}^2\,(10^{-3} \; {\rm eV}^2)$ & $2.5$  & $2.5$        \\
$\D m_{41}^2\,( {\rm eV}^2)$ & $1$  & $1$      \\
$\sin^2{\theta_{12}}$ & $0.32$  & $0.32$      \\
$\sin^2{\theta_{23}}$ & $0.46$  & $0.46$       \\
$\sin^2{\theta_{13}}$ & $0.014$  & $0.014$        \\
\end{tabular}
\end{ruledtabular}
\end{center}
\end{table}

We shall not recount the effective 3+3 formalism that arises from
the eV seesaw model with 4 generations, which we refer to
Ref.~\cite{Hou1}. Our purpose is to assess how the numerics and
conclusions are affected by the new data. To perform our numerical
analysis of a 3+3 picture, we build the $\chi^2$ by using the
three-flavor neutrino oscillation parameters $\D m_{ji}^2$ and
$\sin^2{\theta_{ij}}$ taken from
Ref.~\cite{GonzalezGarcia:2010er}, which are the results of the
global combined analysis done in the framework of the AGSS09 solar
fluxes~\cite{Serenelli:2009yc} and the modified Ga capture
cross-section in Ref.~\cite{Abdurashitov:2009tn}. These are given
explicitly in Table I. We also include the projected value of
$P(\bar{\nu}_\mu \rightarrow \bar{\nu}_e)=0.2334\%$~\cite{Thomas},
and require $\D m^2_{41} \sim 1 \, {\rm eV}^2$ as in the previous
analysis. The seven inputs are listed in the second column in
Table II.

As there are many more parameters, one cannot make a proper fit,
so we just seek to minimize the $\chi^2$ by varying the model
parameters. In this way, we obtain the values for the twelve
parameters $r_{ij}$, $\ep_i$ and $s_i$ (see Ref.~\cite{Hou1} for
definition) of the model at the best fit point, which are given in
Table III. These would be further illustrated below. We then use
these parameter values to get the expectation values for the seven
observables, which are listed in the third column in Table~II. For
comparison, the results from the 2006 study~\cite{Hou1} is given
in Table~IV.

\begin{table}[t]
\caption{Parameter values of the best fit point,
 where the parameters are defined in Ref.~\cite{Hou1}.}
\begin{center}
\begin{ruledtabular}
\begin{tabular}{ccccccc}
Parameter & $\ep_1$ & $\ep_2$ & $\ep_3$ & $s_1$ & $s_2$ & $s_3$ \\
BEST FIT & $0.049$  & $0.018$  & $0.039$ & $0.102$  & $0.9999$ & $-0.72$
\\ \hline
Parameter & $r_{11}$ & $r_{12}$ & $r_{13}$ & $r_{22}$ & $r_{23}$ & $r_{33}$ \\
BEST FIT & $1.263$  & $0.852$  & $1.042$ &  $1.187$ & $0.742$  & $1.223$\\
\end{tabular}
\end{ruledtabular}
\end{center}
\end{table}

\begin{table}[b]
\caption{Inputs~\cite{Lsnd1,Maltoni2} and the
 best fit values from in 2006~\cite{Hou1}, to be compared with Table II.}
\begin{center}
\begin{ruledtabular}
\begin{tabular}{ccc}
 & INPUT & 2006
\\ \hline \\
$P(\bar{\nu}_\mu \rightarrow \bar{\nu}_e)(\%)$ & $0.264$  & $0.15$      \\
$\D m_{21}^2\,(10^{-5} \; {\rm eV}^2)$ & $8.1$  & $8.1$      \\
$\D m_{31}^2\,(10^{-3} \; {\rm eV}^2)$ & $2.2$  & $2.3$        \\
$\D m_{41}^2\,( {\rm eV}^2)$ & $1$  & $1$      \\
$\sin^2{\theta_{12}}$ & $0.30$  & $0.30$      \\
$\sin^2{\theta_{23}}$ & $0.50$  & $0.52$       \\
$\sin^2{\theta_{13}}$ & $0.0$  & $0.0018$        \\
\end{tabular}
\end{ruledtabular}
\end{center}
\end{table}

Comparing these two tables, we can see that the difference between best fit values
and expected (input) values of $P(\bar{\nu}_\mu \rightarrow \bar{\nu}_e)$ in Table II
is smaller than the difference in Table IV,
not just for $P(\bar{\nu}_\mu \rightarrow \bar{\nu}_e)$,
but also for, e.g. $\sin^2{\theta_{23}}$. Comparing the $\chi^2$, we find
\begin{equation}
\chi^{2}_{\rm min,2006}-\chi^{2}_{\rm min,2010} = 1.83,
 \label{eq1}
\end{equation}
where $\chi^{2}_{\rm min,2006}$ and $\chi^{2}_{\rm min,2010}$ are
the $\chi^2$ minima for best fit values in 2006 and 2010. One
could say that the agreement between the scenario of eV seesaw
with four generations and the experimental data has improved.

\subsection{In Search of Sizable \boldmath $\sin^2{\theta_{13}}$}

The plausible value for  $\sin^2{\theta_{13}}$,
the target of Daya Bay and T2K experimemnts, is of great interest.
To discuss this, as in Ref.~\cite{Hou1},
we restrict the $\chi^2$ to the four inputs of
$\sin^2{\theta_{ij}}$ and $P(\bar{\nu}_\mu \rightarrow \bar{\nu}_e)$.
With $\ep_i$ and $r_{ij}$ given as in Table III, we first
fix $s_1=0.1$, the best fit value, and perform a $\chi^2$ fit vs $s_2$ and $s_3$,
then iterate with fixing $s_2 = 0.9999$ ($s_3 = -0.72$) and minimize
$\chi^2$ vs $s_1$ and $s_3$ ($s_1$ and $s_2$).
We find for both cases of fixing $s_1$ or $s_3$ to the best fit values,
$\sin^2{\theta_{23}}$ is quite strongly dependent on $s_2$, and
the value around 0.9999 is preferred. We thus illustrate with
$s_2$ held fixed to this value.

\begin{figure}[t]
  \centering
  \includegraphics[width=0.4\textwidth]{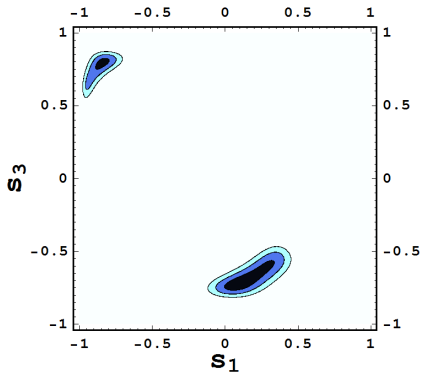}
  \vskip 0.5cm
  \caption{
  Contour-plot of $\chi^2$ vs the mixing angles $s_1$ and
  $s_3$, with $\ep_i$ and $r_{ij}$ as in
  Table IV, and $s_2 = 0.9999$ held fixed.
  The regions in different shades are only indicative, and should
  not be interpreted as the $1\s$, $2\s$ and $3\s$ regions,
  as the rest of the parameters are fixed at the best fit values.
  }
  \label{fig:chiSQ23}
\end{figure}

In Fig. 1 we show the $\chi^2$ contour plot vs $s_1, s_3$. The
three different shaded regions should not be interpreted as the
$1\s$, $2\s$ and $3\s$ regions, since we have fixed the rest of
the parameters to the best fit values. But they do give an
indication of variations around the best fit region under the
above assumptions. The gross features are not too different
from the 2006 study.
For the lower right solution,
$s_1$ has moved closer to zero,
and the strength of $s_3$ has also weakened a little.
For the upper left solution,
$s_3$ is little changed, but $s_1$ has strengthened considerably,
moving the solution more into the corner.

\begin{figure}[b]
  \centering
  \includegraphics[width=0.45\textwidth,height=0.45\textwidth]{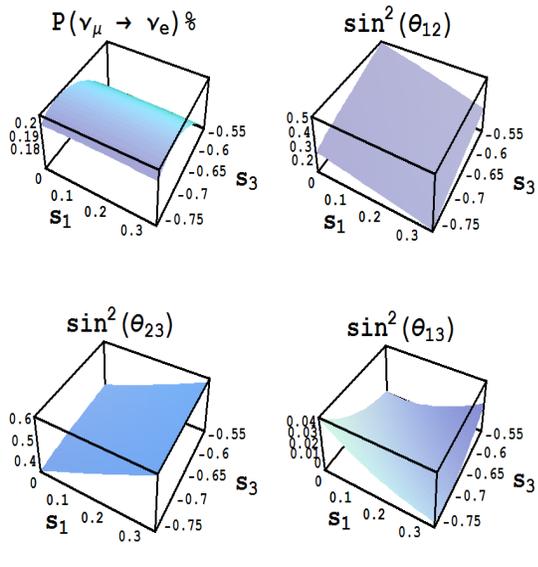}
  \caption{
  $P(\bar{\nu}_\mu \rightarrow \bar{\nu}_e)$, $\sin^2{\theta_{12}}$,
  $\sin^2{\theta_{23}}$ and $\sin^2{\theta_{13}}$ vs $s_1$ and $s_3$,
  corresponding to the lower right solution in Fig. 1, with
  $\ep_i$ and $r_{ij}$ fixed as in Table III, and $s_2$ fixed at 0.9999.}
  \label{fig:allplot1}
\end{figure}

\begin{figure}[b]
  \centering
  \includegraphics[width=0.45\textwidth,height=0.45\textwidth]{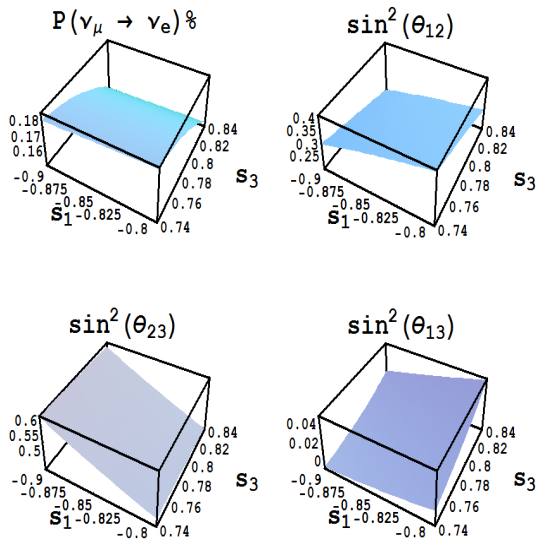}
  \caption{
  Same as Fig. 2, but corresponding to the upper left solution in Fig. 1.}
  \label{fig:allplot2}
\end{figure}

We plot in Fig. 2 the four quantities $P(\bar{\nu}_\mu \rightarrow
\bar{\nu}_e)$, $\sin^2{\theta_{12}}$, $\sin^2{\theta_{23}}$ and
$\sin^2{\theta_{13}}$ vs $s_1$ and $s_3$, for the solution on the
lower right of Fig. 1. The same is plotted in Fig.~3 for the upper
left solution. Again, $\ep_i$ and $r_{ij}$ are fixed as in
Table III, and $s_2$ is held fixed at 0.9999.
The dependence on $s_1$--$s_3$, or shape,
for $P(\bar{\nu}_\mu \rightarrow \bar{\nu}_e)$ differs somewhat from the 2006 case,
with the lower right solution now allowing larger values.
We see that $P(\bar{\nu}_\mu \rightarrow \bar{\nu}_e)$ can reach $ 0.2 \%$
and $\sin^2{\theta_{12}}$ is well within range, and likewise for $\sin^2{\theta_{23}}$.
However, to push $\sin^2{\theta_{13}}$ beyond 0.025,
$\sin^2{\theta_{23}}$ would start to wander away from maximal mixing of
0.5, and values at $\sim 0.4$ or 0.6 has to be tolerated.
The situation is not much changed from 2006, but we note
that the slightly more promising situation for $\sin^2{\theta_{13}}$
can be traced to the more positive input value in Table II,
which now does include SBL data.

\subsection{Neutrino Mass Hierarchy}

The sign of $\D m_{31}^2$ is undetermined by present data (e.g. MINOS experiment).
That is to say, the neutrino mass hierarchy is not known at this time.
We use eV seesaw with four generations scenario to see if we can
gain any access to the neutrino mass hierarchy.
For normal mass hierarchy, the sign of $\D m_{31}^2$ is positive and the best fit value
is $2.47\times10^{-3} \; {\rm eV}^2$, and the numerical analysis has been performed
in the previous section.
For inverted mass hierarchy, the sign of $\D m_{31}^2$ is negative with
best fit value $-2.36\times10^{-3} \; {\rm eV}^2$.
To perform the numerical analysis for inverted mass hierarchy,
we use the best fit values given as in Table III but only replacing that of
$\D m_{31}^2$ by $-2.36\times10^{-3} \; {\rm eV}^2$ to work out $\chi^2$ minimum.
Comparing with the $\chi^2$ minimum for normal mass hierarchy, we find
\begin{equation}
\chi^{2}_{\rm min,inverted}-\chi^{2}_{\rm min,normal} = 1.74,
 \label{eq2}
\end{equation}
in obvious notation. It seems that the agreement of eV seesaw with
four generations scenario with normal mass hierarchy is slightly
better than the agreement with inverted mass hierarchy. We caution
that our study is not a true fit, as there are too many model
parameters.

\section{CP VIOLATION}

So far, we have not considered CP violation effect, and
all phases in the leptonic mixing matrix were set to zero for simplicity,
just as in the previous paper~\cite{Hou1}.
As defined in Ref.~\cite{Hou1}, $U'$ and $U''$ are the rotation matrices
which diagonalize the neutrino and charged-lepton mass matrices, respectively,
and the full leptonic mixing matrix is  $U = U''U'$. Although introducing
CP violation phase to the leptonic mixing matrix adds further to
the already many parameters, there is some motivation to study CP violation
effect, in part because there are both neutrino and antineutrino oscillation data,
and they appear to be somewhat different. We shall therefore give two examples for illustration.

\subsection{Refit with CP Violation Phase}

As charged lepton masses are generated by the Higgs mechanism,
similar to quark masses, we take the left sector rotation matrix
$U''$ to be the usual form
\begin{equation}
{\scriptsize U^{\prime \prime}=\left(
\begin{array}{cccccc}
c_1 c_3 & s_1 c_3 & s_3 e^{-i \delta}& 0 & 0 & 0 \\
-s_1 c_2 - c_1 s_2 s_3 e^{i \delta} & c_1 c_2 - s_1 s_2 s_3 e^{i \delta} & s_2 c_3 & 0 & 0 & 0  \\
s_1 s_2 - c_1 c_2 s_3 e^{i \delta}& -c_1 s_2 - s_1 c_2 s_3 e^{i \delta} & c_2 c_3 & 0 & 0 & 0 \\
0 & 0 & 0 & 1 & 0 & 0 \\
0 & 0 & 0 & 0 & 1 & 0 \\
0 & 0 & 0 & 0 & 0 & 1 \\
\end{array} \right)},
\label{Uprpr}
\end{equation}
where $\delta$ is the CP violation phase and placed in the $13$ element of $U''$.
Because of the presence of $\delta$,
the formulas for $P(\bar{\nu}_\mu \rightarrow\bar{\nu}_e)$
would no longer be equal to that of $P(\nu_\mu \rightarrow \nu_e)$.
We use this new $U''$ matrix with CP violation phase $\delta$,
and the best fit values given as in Table II for inputs,
and redo our numerical analysis. After minimization,
we get the new $\chi^2$ minimum, i.e. $\chi^{2}_{\rm min,CP}$.
We find that introducing CP phase $\delta$
leads to the relative improvement of the fit of
\begin{equation}
\chi^{2}_{\rm min,2010}-\chi^{2}_{\rm min,CP} = 0.004,
 \label{eq4}
\end{equation}
which is negligible. The quality of the global fit does not
improve at all by introducing CP violation phase.
But this is somewhat trivial, since Table II tacitly assumes CP conservation.

\begin{table}[t!]
\caption{Same as Table II, but adding a null $\nu_\mu \rightarrow \nu_e$ oscillation probability
as input, together with the expectation values from the minimization process.}
\begin{center}
\begin{ruledtabular}
\begin{tabular}{ccc}
 & INPUT &
\\ \hline \\
$P(\bar{\nu}_\mu \rightarrow \bar{\nu}_e)(\%)$ & $0.2334$  & $0.1423$      \\
$P(\nu_\mu \rightarrow \nu_e)(\%)$ & $0$  & $0.048$      \\
$\D m_{21}^2\,(10^{-5} \; {\rm eV}^2)$ & $7.59$  & $7.59$      \\
$\D m_{31}^2\,(10^{-3} \; {\rm eV}^2)$ & $2.5$  & $2.5$        \\
$\D m_{41}^2\,( {\rm eV}^2)$ & $1$  & $1$      \\
$\sin^2{\theta_{12}}$ & $0.32$  & $0.32$      \\
$\sin^2{\theta_{23}}$ & $0.46$  & $0.43$       \\
$\sin^2{\theta_{13}}$ & $0.014$  & $0.017$        \\
\end{tabular}
\end{ruledtabular}
\end{center}
\end{table}

\subsection{An Extreme Example of CP Violation}

The updated MiniBooNE $\nu_\mu \rightarrow \nu_e$ result indicates
an excess of $\nu_e$ events at low energy, but
no $\nu_\mu \rightarrow \nu_e$ oscillations at the $L/E \sim 1 \, {\rm m/MeV}$
of the LSND excess, which is for antineutrinos.
Furthermore, neither NUMI~\cite{Adamson} nor NOMAD~\cite{NOMAD}
found evidence for $\nu_\mu \rightarrow \nu_e$ oscillations. Thus,
we consider $P(\nu_\mu \rightarrow \nu_e) = 0\%$ at $L/E = 1 \,
{\rm m/MeV}$ as a new additional input, to illustrate an extreme
situation where there is a striking difference between the
behavior of $\nu_\mu \rightarrow \nu_e$ and $\bar{\nu}_\mu
\rightarrow\bar{\nu}_e$ oscillations. We stress that this is only
an illustration, since neither the LSND effect for $\bar{\nu}_\mu
\rightarrow\bar{\nu}_e$ oscillations, nor the absence of $\nu_\mu
\rightarrow \nu_e$, are firmly established. Furthermore, we
clearly have too many parameters, hence we are just performing
$\chi^2$ illustrations, not a fit.

The new ad hoc input, together with the 7 inputs of Table II, are
now listed in Table V, with which we build the new $\chi^2$ in the
framework with CP violation. After minimizing the $\chi^2$, we use
the thirteen parameter values of the best fit point to get the the
expectation values for the eight observables, which are listed in
the third column in Table V. There is now a slight tension in
$\sin^2\theta_{23}$ (though allowed), while neither the
antineutrino, nor the neutrino oscillation probabilities can reach
the input values, but this may be reasonable. The best fit value
of CP violation phase $\delta$ is $0.37\pi$, or of order
$66^\circ$, which is of course rather large. Note also that the
$\sin^2{\theta_{13}}$ value is now larger than in Table II, which
is again reasonable, since a finite $\sin^2{\theta_{13}}$ is a
prerequisite for CP violation. This illustrates the efficacy of
the Daya Bay and T2K experiments, as well as a future CPV program.

\section{Summary}

We have updated the numerics of a 3+3 active/sterile neutrino picture,
which arises naturally from a four lepton generation model with eV seesaw,
i.e. with right-handed Majorana mass scale $M_R$ at eV order. The major new data are
the MiniBooNE results of the past few years, which do not confirm the LSND
$\bar{\nu}_\mu \rightarrow\bar{\nu}_e$ result. Since the 3+3 model has too
many parameters, while data is more or less consistent with a 3+1 picture,
we take global fit results using the latter framework as input, and use
a $\chi^2$ minimization to find the best values for neutrino mass differences,
mixing angles, and $\bar{\nu}_\mu \rightarrow\bar{\nu}_e$ probability.
Not unexpectedly, the consistency of the 3+3 picture with current data
improves over the consistency with data in 2006. We continue to find
it difficult for $\sin^2{\theta_{13}}$ to be larger than 2.5\%, while
finding a slightly better $\chi^2$ for normal mass hierarchy over the
inverted mass hierarchy. These, however, could be artefacts of not being
able to do a real fit.

If one continues to take the LSND $\bar{\nu}_\mu \rightarrow\bar{\nu}_e$ result
seriously (which provides the eV scale), together with the consensus of null results for
$\nu_\mu \rightarrow \nu_e$ oscillations, we have put in the CP violation phase
in the charged lepton mixing matrix, and illustrated the possibility of CP violation.
The world of large CP violation effect in $\bar{\nu}_\mu \rightarrow\bar{\nu}_e$
vs $\nu_\mu \rightarrow \nu_e$ oscillations, with large CPV phase
and finite $\sin^2{\theta_{13}}$, could be a striking one.

\vskip 0.3cm \noindent{\bf Note Added}.\ Our Table I uses version
2 of Ref.~\cite{GonzalezGarcia:2010er}, which is the published
one. Subsequently, these authors have posted version 3 on the
arXiv, which includes new results on $\nu_e$ appearance from
MINOS, which leads to a smaller $\sin^2{\theta_{13}}$ (best fit
at 0.008). However, we have discussed how things will be for a
lower $\sin^2{\theta_{13}}$, so we leave the paper as is.

\vskip 0.3cm  \noindent {\textbf{Acknowledgement}.}\ \ We thank
T.~Schwetz and G.~Karagiorgi for useful communications. The work
of WSH is supported in part by NSC97-2112-M-002-004-MY3 and
NTU-98R0066. The work of FFL is supported by NSC98-2811-M-002-102.

\end{document}